\documentclass[journal=jacsat,manuscript=article]{achemso}

\usepackage[version=3]{mhchem} 

\author{Vasileios Fotopoulos}
\email{vasileios.fotis.19@ucl.ac.uk}
\affiliation[University College London]
{Department of Physics and Astronomy, University College London, Gower Street, London WC1E 6BT, United Kingdom}
\author{Jack Strand}
\affiliation[University College London]
{Department of Physics and Astronomy, University College London, Gower Street, London WC1E 6BT, United Kingdom}
\alsoaffiliation[Nanolayers]{Nanolayers Research Computing Ltd., London, UK}
\author{Manuel Petersmann}
\affiliation[KAI]{
	 KAI--Kompetenzzentrum Automobil- und Industrieelektronik GmbH, Europastrasse 8, 9524 Villach, Austria}
  \author{Alexander L. Shluger}
\affiliation[University College London]
{Department of Physics and Astronomy, University College London, Gower Street, London WC1E 6BT, United Kingdom}

\title[An \textsf{achemso} demo]
  {First principles study on the segregation of metallic solutes and non-metallic impurities in\\ Cu grain boundary}

\abbreviations{IR,NMR,UV}
\keywords{American Chemical Society, \LaTeX}

\begin{document}


\begin{abstract}
Metallic dopants have the potential to increase the mechanical strength of polycrystalline metals. These elements are expected to aggregate in regions of lower coordination, such as grain boundaries. At the grain boundaries, they can have a beneficial (toughening) or detrimental effect (e.g. grain boundary embrittlement). In this study, we employ Density Functional Theory (DFT) to compute the segregation energies of various metallic and other non-metallic elements to determine their effect when introduced in a symmetric Cu grain boundary. The study results may be used to qualitatively rank the beneficial effect of certain metallic elements, such as V, Zr, and Ag, as well as the strong weakening effect of non-metallic impurities like O, S, F and P. Furthermore, the induced local distortion is found to be proportional to the weakening effect of the elements.
\end{abstract}


\section{Introduction}\label{sec1}

Degradation phenomena in metals have been reported to initiate at certain microstructural features, such as grain boundaries (GBs)~\cite{moser2021electropolishing,konishi2002effect,hallberg2018crystal,stopka2020microstructure}. Experimental results have illustrated how the level of corrosion and degradation of metals under high-stress conditions depends on the concentration of certain elements~\cite{hu2018role}. Impurity-induced embrittlement accounts for many cases of brittle failure of polycrystalline metals~\cite{duscher2004bismuth,ludwig2005situ,laporte2009intermediate}. Impurity-induced embrittlement is commonly attributed to a chemical effect of the segregation of solutes, which is believed to change the bonding strength at GBs~\cite{duscher2004bismuth,kang2013origin}, or to a size (strain) effect that is associated with the size mismatch between the solute and host atoms~\cite{schweinfest2004bismuth,lozovoi2006structural}. The segregation of metallic dopants and non-metallic impurity elements from GBs in metals plays a key role in the design of novel materials~\cite{scheiber2021segregation}. However, a clear correlation between the properties of solutes and the strengthening or weakening effect that they can have when introduced in the GBs of metals is still missing. 

 In light of recent experimental studies in bicrystalline Fe, solutes that decorate grain boundaries significantly affect their chemical composition, charge distribution, structural properties, and therefore their mechanical stability~\cite{zhou2023atomic}. GBs in polycrystalline metals can be engineered and decorated with certain elements to increase their resistance against corrosion effects. Thus, identifying the elements that could potentially be employed as GB decorative solutes is of great interest. First-principles simulations on 3d transition metal solutes in Fe~\cite{xu2019grain,mai2022segregation}, Au~\cite{scheiber2021segregation}, Mo~\cite{scheiber2018impact} and Ni~\cite{razumovskiy2015first} illustrated that their presence at GBs can significantly increase their tensile strength and resistance against embrittlement phenomena. Cu, a metal used in a wide range of applications, including electronic devices, showed prominent degradation effects under thermal cycling~\cite{moser2019novel,moser2021electropolishing}. Such phenomena often initiate at crystal grain boundaries and can have devastating effects on the performance of these devices. Doping of grain boundaries with metallic elements can be an efficient solution to this problem~\cite{huang2018uncovering}. However, although several properties have been proposed to play a critical role in the weakening or strengthening effect that these elements can have, a correlation between the above-mentioned effects and the elements' induced local distortion into the lattice hasn't been established.

This work complements previous theoretical studies of metallic segregants in Cu GBs. Density functional theory (DFT) simulations are conducted to determine the most favorable segregation sites of 3d transition and other metals in a symmetric Cu GB, along with their local strengthening or weakening effect. We highlight the beneficial effects of V, Zr, and Ag. Finally, we show how the computed segregation energies are linked with the solutes' induced lattice relaxation. Metallic solutes with low GB segregation energies induce small relaxation. On the other hand, non-metallic impurities with a high computed strengthening energy, such as O, S, F, and P, cause severe local displacements of Cu atoms. Such findings highlight the possibility of decorating Cu GBs with specific metallic elements to reduce the degradation effects in Cu.

\section{\label{sec:level1}Methods}

\begin{figure}[htp]
    \centering
    \includegraphics[scale=0.55]{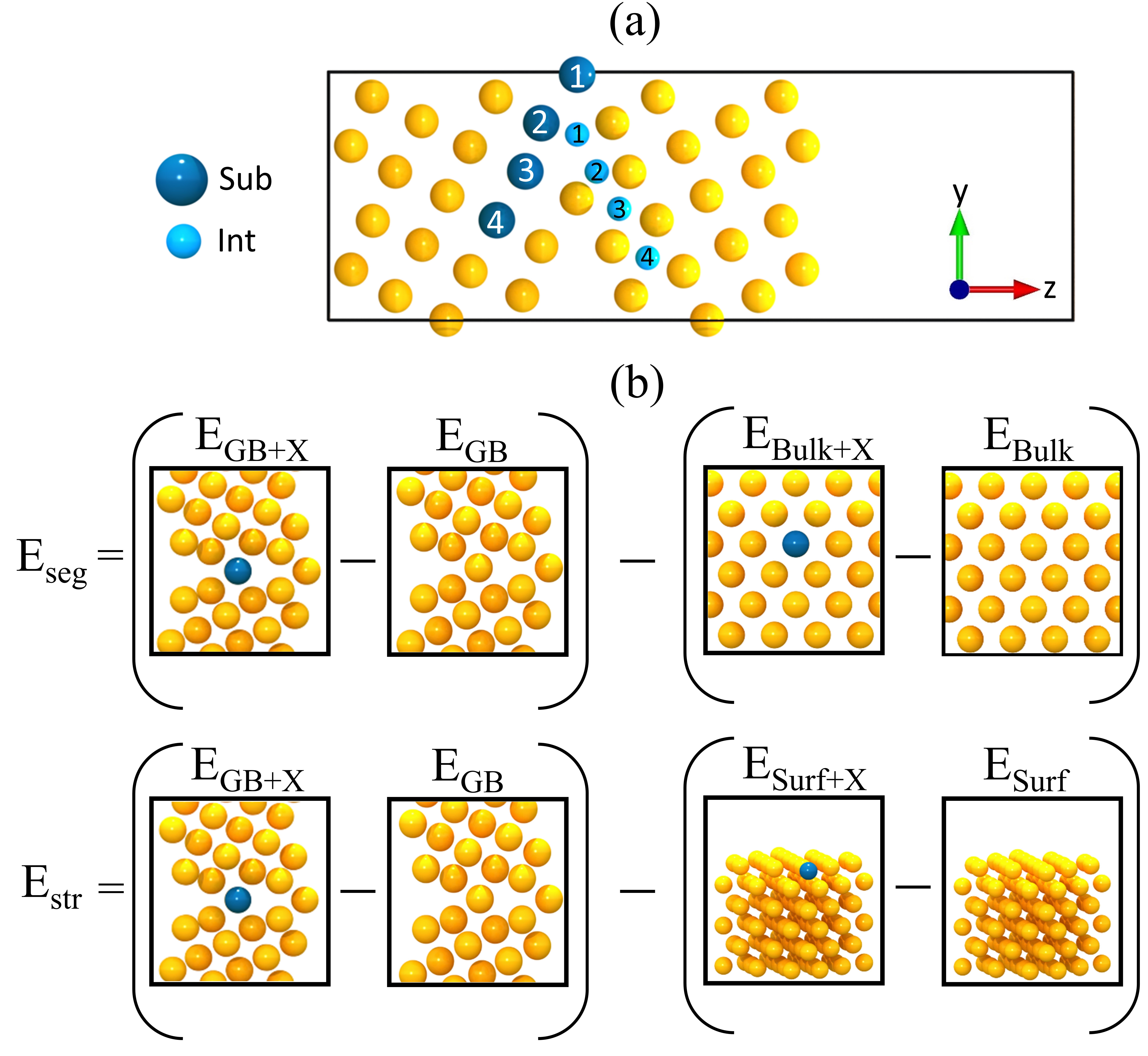}
    \caption{(a) Examined substitutional and interstitial sites of dopants/impurities in (210)[100]$\Sigma5$ Cu grain boundaries. (b) Schematics illustrating (top) the segregation energy and (bottom) strengthening energies of substitutional impurities/dopants in the grain boundaries and triple points, respectively.}
    \label{fig:1}
\end{figure}

DFT calculations are performed using the Vienna Ab Initio Simulation Package (VASP) \cite{kresse1993ab, kresse1996efficient, kresse1996efficiency}, with the GGA exchange-correlation functional Perdew-Burke-Ernzerhof (PBE) \cite{perdew1996generalized}. For the investigation of grain boundaries and bulk properties of copper, 76-atom and 108-atom periodic cells are utilized, respectively. The specific grain boundary chosen for examination is the (210)[100]$\Sigma5$ twin boundary (shown in Figure \ref{fig:1}(a)), due to its high symmetry and low-energy characteristics \cite{wu2016first}. Consistent with previous studies \cite{nazarov2012vacancy, nazarov2014ab}, the calculations employed converged 5$\times$4$\times$1 and 4$\times$4$\times$4 k-point grids for the 76 and 108-atom simulation cells, respectively. An energy cut-off of 450\,eV is used, as established in previous Cu simulations \cite{fotopoulos2023thermodynamic}.

Segregation energies were calculated using the formula:

\begin{equation}
    E_{seg}=(E_{GB+X}-E_{GB})-(E_{Bulk+X}-E_{Bulk}),
    \label{eq:impurities}
    \end{equation} 
\noindent{where} $E_{GB+X}$ and $E_{GB}$ are the energies of GB cells with and without segregants (X) and $E_{Bulk+X}$, $E_{Bulk}$ the energies of bulk cells with and without segregants, respectively. Negative and positive segregation energies illustrate favorable GB segregation and antisegregation, respectively. For the determination of the segregation energies related to interstitial impurities, octahedral and tetrahedral interstitial sites are considered in the bulk. Figure \ref{fig:1}(a) presents the four interstitial and substitutional positions considered at the GB for each segregant. 

The effect of impurities on the strength of the grain boundary can be derived from the strengthening energy. This parameter provides insight into whether the solutes exhibit a preference for locating themselves at grain boundaries as opposed to surface positions. The calculation of DFT strengthening energies is carried out according to the following expression:

\begin{equation}
    E_{str}=(E_{GB+X}-E_{GB})-(E_{Sur+X}-E_{Sur}),
    \label{eq:strength}
    \end{equation} 

\noindent{where} E$_{Sur}$ and E$_{Sur+X}$ symbolize the energies associated with the pure Cu surface and the Cu surface simulation cells containing one solute atom, respectively. For the surface simulations, 108-atom simulations (100) slab models are used with a thickness of 10.86\,{\AA}. The strengthening energy, when negative, indicates an enhancement of the grain boundary strength due to the impurity, while a positive value implies a weakening effect. Schematic representations of the segregation and strengthening energies are shown in Figure \ref{fig:1}(b).

\section{\label{sec:level1}Results}

\subsection{Segregation Energies}

\begin{figure}
 \centering
   \includegraphics[scale=0.5]{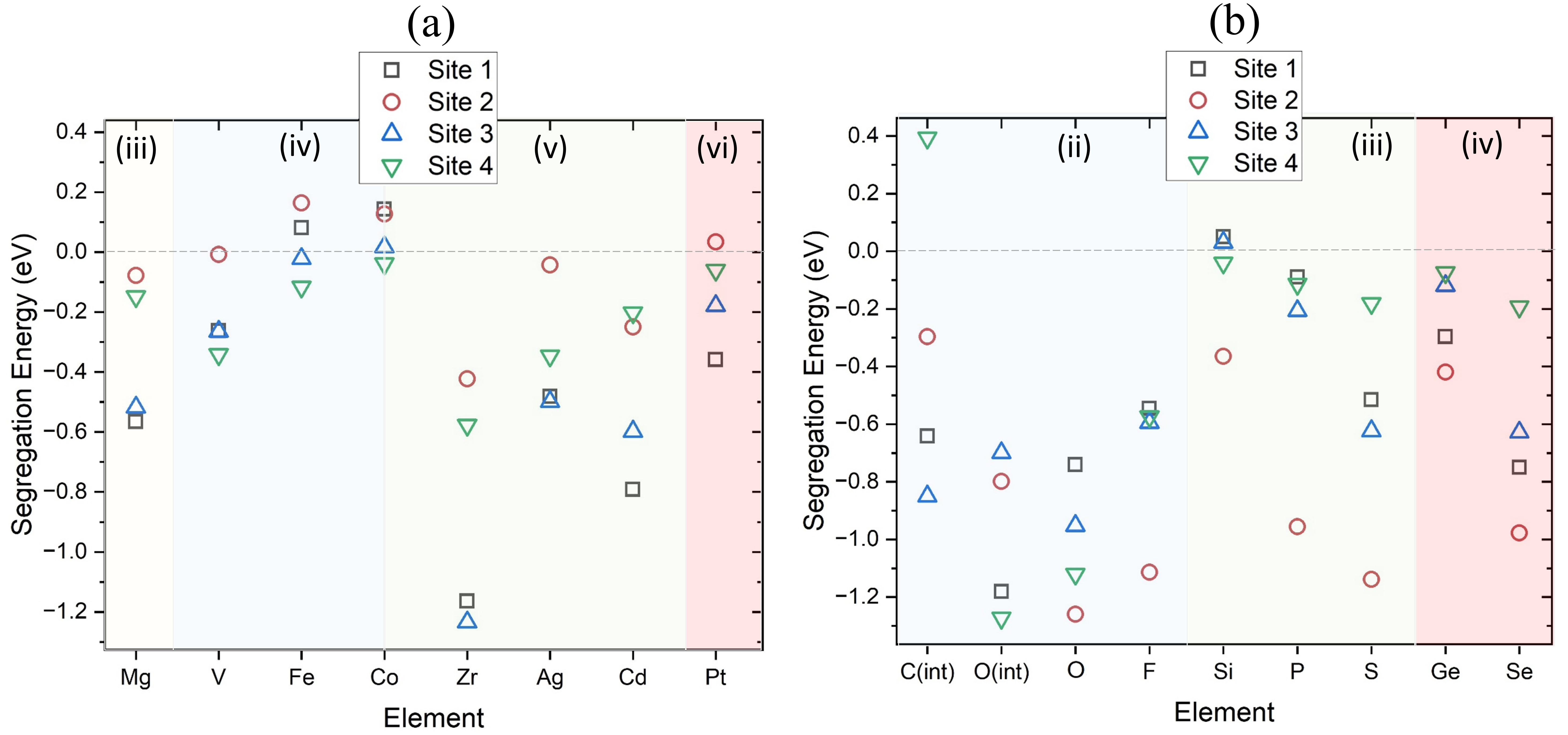}\\
\caption{Computed segregation energies for (a) metallic and (b) non-metallic elements for different sites. Different colored areas correspond to elements of different groups. Negative energies correspond to favorable segregation at the GB, whereas positive energies correspond to antisegregation. The int represents interstitial atoms, whereas the rest of the elements correspond to substitutional atoms.}
    \label{fig:2}
\end{figure}

To determine the most favorable sites, as shown in Figure \ref{fig:1}(a), of the impurities examined, the segregation energies of all elements are calculated. Figure \ref{fig:2}(a) includes all the computed segregation energies for the various sites. The plots are divided into colored regions on the basis of the group of elements on the periodic table. As seen in the figure, most of the metallic dopants are found to occupy most preferably substitutional sites 1 and 3. Fe and Co dopants show segregation energies close to 0\,eV, indicating that these elements do not show an energetic preference to be at the GB instead of the bulk region. 

For non-metallic impurities, our results show that all substitutional elements will occupy site 2. Also, non-metallic elements show a significantly stronger tendency to segregate at the GB compared to metallic solutes, with F, S, O, and P resulting in the lowest segregation energies. C is the only element that shows a strong preference to segregate as an interstitial atom. On the basis of these results, elements such as C, O, F, P, S, and Se, when introduced in polycrystalline Cu, are expected to be at considerably higher concentrations at the GBs of the crystal.

\begin{table}
\centering
\footnotesize

\caption{Segregation energies comparison between the results from the current work and previous DFT studies for metallic and non-metallic elements. Values with $^*$ and $^{**}$ correspond to segregation energies computed in (221)[110]$\Sigma$9 and (310)[001]$\Sigma$5 GB, respectively. The rest of the values, including our results, were obtained using (210)[100]$\Sigma$5 GB simulations cells.} 
\begin{tabular}{llllllllll}

 \multicolumn{4}{c}{}\\ 
   \\  \hline
\cline{1-6} 
        & Element & Reference & E$_{seg}$\,(eV) &\vline E$_{seg}$\,(eV) [Our work] &
        \\  \hline\hline
  & C (int) & Wurmshuber \cite{wurmshuber2022mechanical} & -0.8 &\vline\hspace{0.2cm} -0.85 & \\
& C (int) & Huang (2018) \cite{huang2018uncovering} & -0.9 &\vline\hspace{0.2cm} -0.85& \\  
& O (int) & Bodlos \cite{bodlos2023modification} & -1.2 &\vline\hspace{0.2cm} -1.27  & \\
  & O & Bodlos \cite{bodlos2023modification} & -1.15 &\vline\hspace{0.2cm} -1.26  & \\
& Si & Huang (2018) \cite{huang2018uncovering} & -0.3 &\vline\hspace{0.2cm} -0.36 & \\
& Si & Li \cite{li2017impurity} & -0.58$^{**}$ &\vline\hspace{0.2cm} -0.36 & \\
   & P & Li \cite{li2017impurity} & -0.95 &\vline\hspace{0.2cm} -0.9 & \\
      & P & Lousada \cite{lousada2020segregation} & -0.65$^*$ &\vline\hspace{0.2cm} -0.9 & \\
   & S & Bodlos \cite{bodlos2023modification} & -1.2 &\vline\hspace{0.2cm} -1.14 & \\
   & S & Wang \cite{wang2018first} & -1 &\vline\hspace{0.2cm} -1.14 & \\
& S & Li \cite{li2017impurity} & -1.15$^{**}$ &\vline\hspace{0.2cm} -1.14 & \\
      & S & Lousada \cite{lousada2020segregation} & -0.75$^*$ &\vline\hspace{0.2cm} -1.14 & \\
& Ge & Razumovskiyy \cite{razumovskiy2018solute} & -0.3 &\vline\hspace{0.2cm} -0.42 & \\
 & Se & Razumovskiyy \cite{razumovskiy2018solute} & -0.8 &\vline\hspace{0.2cm} -0.98 & \\
  \hline
   & Mg & Huang (2020)~\cite{huang2020understanding} &-1$^{**}$ &\vline\hspace{0.2cm} -0.57 & \\
    & V & Huang (2020)~\cite{huang2020understanding} &-0.2$^{**}$ &\vline\hspace{0.2cm} -0.34 & \\
  & Co & Razumovskiyy \cite{razumovskiy2018solute} &\hspace{0.0cm} 0 &\vline\hspace{0.2cm} -0.04 & \\
  & Ag & Razumovskiyy \cite{razumovskiy2018solute} & -0.6 &\vline\hspace{0.2cm} -0.5 &\\
  & Ag & Huang (2019) \cite{huang2019combined} & -0.8 &\vline\hspace{0.2cm} -0.5 &\\
    & Zr & Huang (2018) \cite{huang2018uncovering} & -1.65 &\vline\hspace{0.2cm} -1.23 &\\
 \hline\hline
\end{tabular}
    
\label{tab:1}
\end{table}

Table \ref{tab:1} includes the comparison between the segregation energies calculated in the current work and previous studies. For non-metallic impurities, the computed segregation energies are within 0.2\,eV compared to previous studies. The only exceptions are seen in the case where a different $\Sigma$5 GB ((310)[001]$\Sigma$5 GB~\cite{li2017impurity,huang2020understanding}), or a different $\Sigma$ symmetry ((221)[110]$\Sigma$9~\cite{lousada2020segregation}) is used. For the segregation energies of the metallic dopants, our results agree well with the results of ref.\cite{razumovskiy2018solute}, while the study ref.\cite{huang2018uncovering} showed lower energies by 0.3 and 0.42\,eV for Ag and Zr, respectively. In general, the table shows that there is a clear trend for certain non-metallic (C, O, P, S, Se) and metallic dopants (Zr) to strongly segregate at the grain boundaries. The rest of the elements show small to close to no preference for being at the grain boundaries instead of the bulk of Cu.

\begin{figure}
 \centering
   \includegraphics[scale=0.5]{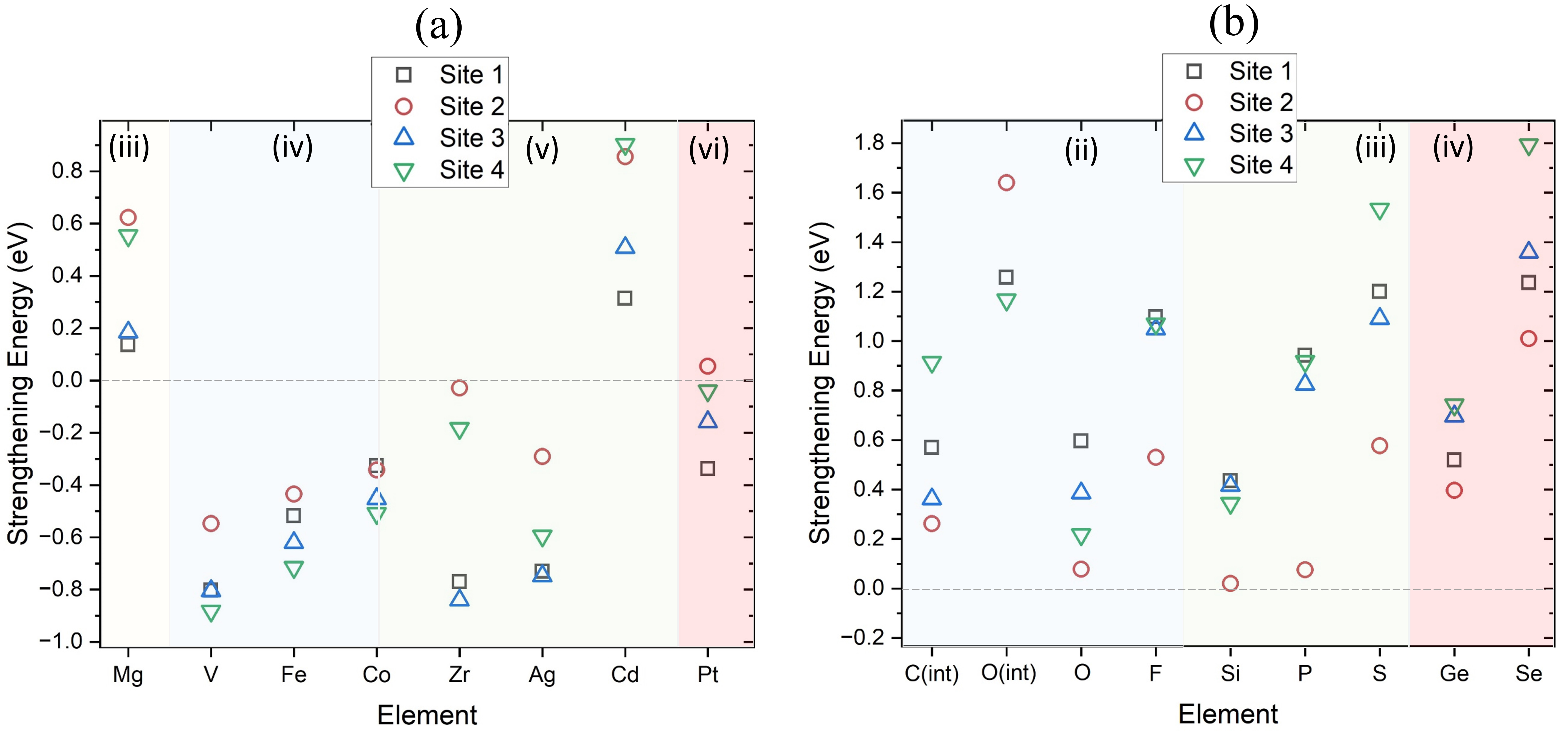}\\
\caption{Computed strengthening energies for (a) metallic and (b) non-metallic elements for different sites. Different colored areas correspond to elements of different groups. Negative strengthening energies correspond to a strengthening effect, whereas positive strengthening energies correspond to a weakening effect.}
    \label{fig:3}
\end{figure}

\subsection{Strengthening Energies}

In Figure \ref{fig:3}, the effect of the studied elements on the strength of the GB is shown when introduced at various sites. Negative strengthening energies correspond to a strengthening effect, whereas positive strengthening energies correspond to a weakening effect. On the basis of these results, one can deduce that most of the metallic solutes show negative strengthening energies (strengthening effect), whereas all non-metallic elements show positive strengthening energies (weakening effect). The latter shows that our findings support previous studies and highlight the differences between metals and non-metals when it comes to their impact on GBs~\cite{razumovskiy2018solute,huang2018uncovering,huang2019combined}. 

Table \ref{tab:2} summarizes the main conclusions of our investigation so far. One can see that all of the examined non-metallic elements are expected to have a weakening effect when they are introduced into the GB of Cu. From the metallic dopants, excluding Mg and Cd, all of the rest showed a strengthening effect. The weakening effect of Mg and Cd can be attributed to their small atomic radii. The grain boundary energy is expected to decrease as the atomic radius of the metallic dopants increases~\cite{huang2019combined}. These results agree with the conclusions drawn in ref.\cite{razumovskiy2018solute}, where metallic dopants with energies close to zero, in our case Co and Fe, are expected to strengthen the GB. Also, in agreement with the aforementioned study, non-metallic impurities relaxing to the site next to the GB plane (site 2) weaken the GB. Furthermore, as highlighted in previous theoretical works~\cite{huang2020understanding}, non-metallic solutes with larger atomic radius segregate more easily at the grain boundary but cause more significant grain boundary embrittlement.

\begin{table}
\centering
\footnotesize

\caption{Properties of the examined non-metallic impurities and metallic solutes along with the identified most favorable sites and their effect. W and S correspond to weakening and strengthening effects, respectively.} 
\begin{tabular}{llllllllll}

 \multicolumn{6}{c}{}\\ 
   \\  \hline
\cline{1-8} 
        & Element & Group & Mass (amu) & Radius (pm) & Site & Effect (S/W) &
        \\  \hline\hline
& C & 2 & 12.01 & 170 & 3 (int) & W &\\
& O & 2 & 15.99 & 152 & 4 (int) & W &\\
& O & 2 & 15.99 & 152 & 2 & W &\\
& F & 2 & 18.99 & 147 & 2 & W &\\
& Si & 3 & 28.09 & 210 & 2 & W &\\
& P & 3 & 30.97 & 195 & 2 & W &\\
& S & 3 & 32.07 & 180 & 2 & W &\\
& Ge & 4 & 72.64 & 211 & 3 & W &\\
& Se & 4 & 78.96 & 190 & 2 & W &\\
\hline
 & Mg & 3 & 24.31 & 173 & 1 & W &\\
& V & 4 & 50.94 & 179 & 4 & S &\\
& Fe & 4 & 55.85 & 126 & 4 & S &\\
& Co & 4 & 58.93 & 200 & 4 & S &\\
& Zr & 5 & 91.22 & 206 & 3 & S &\\
& Ag & 5 & 107.87 & 172 & 1/3 & S &\\
& Cd & 5 & 112.41 & 158 & 1 & W &\\
& Pt & 6 & 175 & 195.08 & 1 & S &\\
\hline
\hline
\end{tabular}
    
\label{tab:2}
\end{table} 

\subsection{Lattice Distortion Effects}

\begin{figure}
 \centering
\includegraphics[scale=0.5]{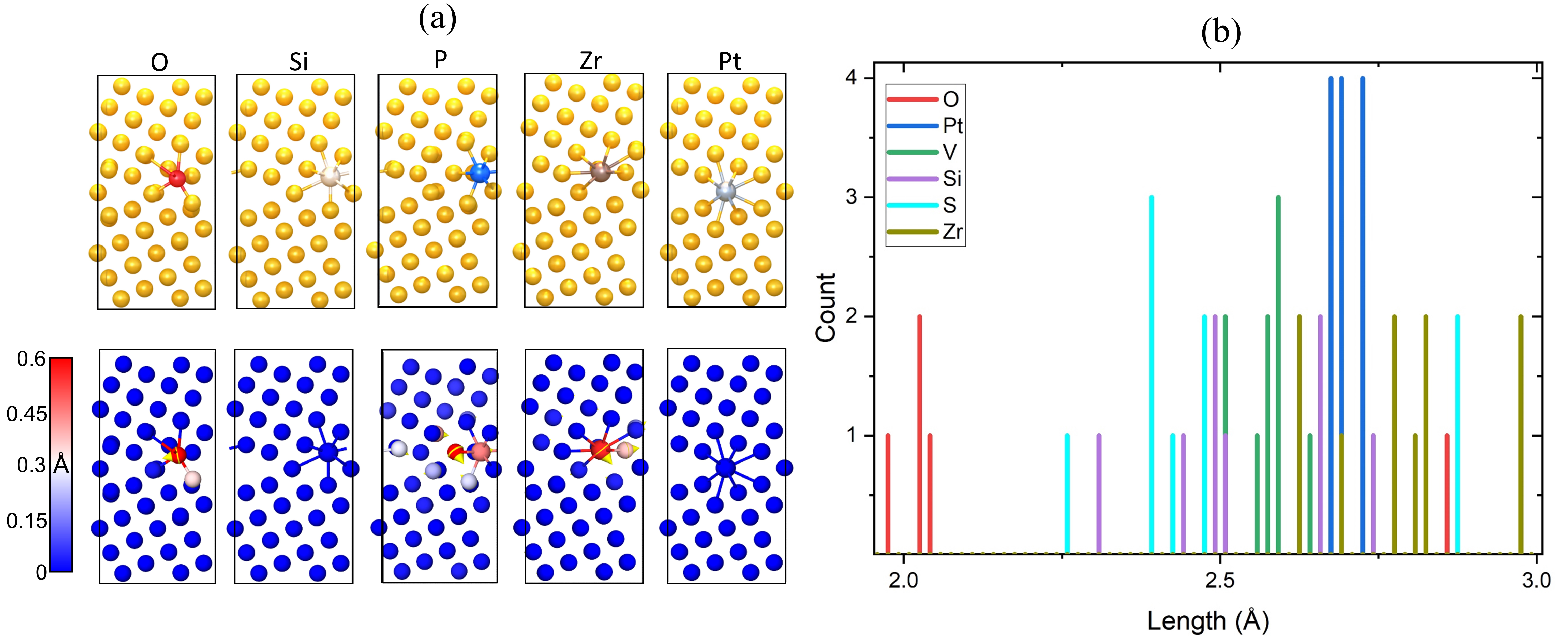}\\
\caption{(a) Fully relaxed configurations along with their displacement magnitudes (bottom) for O, Si, P, Zr, Pt. Yellow arrows correspond to the displacement vectors. For displacements, the configuration prior to relaxation is used as a reference. (b) Bond analysis for the aforementioned elements. Bond analysis illustrates the distribution of distances between the dopants/impurities and nearby Cu atoms within a cut-off distance of 3\,{\AA}. The y-axis corresponds to the number of Cu atoms at a certain distance from the dopant/impurity.}
    \label{fig:4}
\end{figure}

In the previous sections, the segregating and strengthening energies were computed, and the most favorable segregation sites were identified. The computed energetic parameters follow the pattern already identified by previous studies~\cite{razumovskiy2018solute, huang2018uncovering,huang2019combined}. We now move on to examine the effect of impurities/dopants on local lattice distortions. Figure \ref{fig:4}(a) includes the fully relaxed configurations for some of the tested metallic dopants along with non-metallic elements at their most favorable sites. In addition, displacements are also included, both in the form of colored atoms and as displacement vectors. Relaxed configurations for O, Si, P, Zr and Pt are shown in Figure \ref{fig:4}(a). As can be deduced, substitutional metallic solutes cause minimal relaxation. Non-metallic elements, either substitutional or intersitial, induce significant local relaxation, which explains the weakening effect that these elements have. Non-metallic impurities with a small preference to segregate at the GB, like Si, caused negligible local distortion. Figure \ref{fig:4}(b) includes the results of the bond analysis. Bond analysis illustrates the distribution of distances between the dopants/impurities and nearby Cu atoms within a cut-off distance of 3.0\,{\AA}. Any distances beyond the mentioned cut-off are ignored. All metallic dopants showed distances from the adjacent Cu atoms of approximately 2.6\,{\AA}, which is the same as the Cu-Cu distance in the perfect lattice. The latter highlights the minimal distortion that the metallic dopants introduced. On the other hand, non-metallic impurities show peaks at lower than 2.5\,{\AA}.

\section{Conclusions}

We employed DFT to investigate the energetic properties of metallic dopants and non-metallic impurities introduced into the grain boundaries of Cu. Among the metallic dopants, such as Mg, V, Fe, Co, Ag, and Zr, a clear trend towards segregation within the grain boundaries rather than the bulk was found. The aforementioned elements also showed negative strengthening energies, indicating that they are expected to locally increase the resistance of the crystal against decohesion. Furthermore, the results highlighted a preference for non-metallic impurities, like O, P, F, S, and Se, to segregate at the grain boundaries as opposed to the bulk of Cu. This enhanced affinity of impurities to grain boundaries, along with the predicted local weakening effect of these elements, could potentially explain experimental findings that point to the initiation of degradation phenomena in these regions. A link was determined between the segregation energies and the induced relaxation effects. Metallic solutes with segregation energies close to zero were shown to induce minimal distortion, whereas non-metallic elements that strongly segregate at the GB caused severe displacement of nearby Cu atoms.

 Grain boundaries are known to play an important role in the properties of metals. On the basis of our findings, the incorporation of metallic dopants such as V, Zr, and Ag at the grain boundaries of Cu has the potential to improve the mechanical properties of these regions and enhance their resistance against degradation. Further studies are needed to understand the behavior of these impurities when introduced into polycrystalline Cu.

\section*{Acknowledgements}
A.L.S. acknowledges funding by EPSRC (grant EP/P013503/1). V.F. would like to acknowledge funding by EPSRC (grant EP/L015862/1) as part of the CDT in molecular modeling and materials science. Computational resources on ARCHER2 (http://www.archer2.ac.uk) were provided via our membership in the UK's HPC Materials Chemistry Consortium, which is funded by EPSRC (EP/L000202, EP/R029431). V.F. and A.L.S. would like to thank Rishi Bodlos, Lorenz Romaner, and Ernst Kozeschnik for useful comments and help with calculations.
\bibliographystyle{unsrt}
\bibliography{manuscript}

\end{document}